# ZrN nucleation layer provides backside ohmic contact to MBE-grown GaN nanowires

Stanislav Tiagulskyi,*[1] Roman Yatskiv,[1] Marta Sobanska,[2] Karol Olszewski,[2] Zbigniew R. Zytkiewicz,[2] and Jan Grym,*[1]

*e-mail address: tiagulskyi@ufe.cz, grym@ufe.cz

[1]Institute of Photonics and Electronics, Czech Academy of Sciences, Chaberská 57, 18200 Prague, Czech Republic. E-mail: tiagulskyi@ufe.cz, grym@ufe.cz.
[2]Institute of Physics, Polish Academy of Sciences, Al. Lotnikow 32/46, 02 668 Warsaw, Poland.

**Abstract**

Self-assembled GaN nanowires are typically grown on Si substrates with convenient nucleation layers. Light-emitting devices based on arrays of GaN nanowires require that the nucleation layer is electrically conductive and optically nontransparent to prevent the absorption of generated light in the Si substrate. This study reports the molecular beam epitaxial growth of GaN nanowires on ZrN nucleation layers sputtered on sapphire and demonstrates that ZrN provides ohmic contact to dense vertical arrays of n-type GaN nanowires. The ohmic nature of the ZrN/n-type GaN nanowire contact is evidenced by the measurement of the current-voltage characteristics of individual as-grown nanowires using nanomanipulators in a scanning electron microscope. The limitations and advantages of single-nanowire measurements are discussed, and approaches to overcome these limitations are proposed. The feasibility of this concept is demonstrated by the measurement of single NWs with a p-n junction, exhibiting highly rectifying characteristics.

**Introduction**

Gallium Nitride (GaN) is the cornerstone of advanced electronic and optoelectronic devices. GaN exhibits exceptional chemical and thermal stability and high resistance to degradation under extreme environmental conditions. This leads to a long lifetime and high reliability of the GaN-based devices. Tunability of the bandgap of GaN-based alloys and their high photoelectronic efficiency are responsible for the high performance of optoelectronic devices using this class of materials, including light-emitting devices, lasers, and photodetectors [1].

Further improvement of GaN-based nanoscale optoelectronic devices has been achieved through the implementation of GaN nanowires (NWs) [2]. The reduced lateral size of NWs is particularly advantageous for effective doping and for the elimination of extended defects, leading to high crystalline quality and excellent optical properties of NWs. The strain induced by dopant incorporation is effectively relaxed during the growth of NWs [3-5]. Even when grown on lattice-mismatched substrates, NWs exhibit excellent crystal quality; the strain is elastically relaxed, and dislocations are confined to the NW/substrate interface or terminated at the sidewalls of the NWs [5, 6]. NWs exhibit reduced susceptibility to high power consumption and self-heating, rendering them advantageous for energy-efficient optoelectronic devices [7]. Moreover, the light-concentrating property of vertically oriented NWs enhances the light absorption of NW-based devices [8].

GaN NWs grown by molecular beam epitaxy (MBE) are typically prepared on Si substrates, which has both benefits and drawbacks. Si substrates are advantageous because of their compatibility with standard CMOS technology, which facilitates integration with existing microelectronics. However, the use of Si substrates presents challenges in optoelectronic applications. In particular, the lack of transparency in the visible spectrum significantly limits the external efficiency of light-emitting devices and complicates optical access to GaN NWs [9]. Alternatively, promising results for GaN NW growth have been demonstrated on AlN [10] and amorphous buffer layers, such as $SiO_2$ [11], $Al_xO_y$ [12-14], and $SiN_x$ [15]. However, these buffer layers are insulating and require additional steps to obtain electrical contacts, typically with the underlying Si substrate [16-19]. This can hinder the carrier transport and heat dissipation at the interface. Furthermore, owing to the optical transparency of these buffer layers, a significant portion of the light generated in GaN NW LEDs is absorbed by the underlying Si substrate, leading to light loss and reduced device efficiency. One potential solution to this issue is to grow GaN NWs directly on metal substrates. Metal buffer layers show high electrical and thermal conductivities and optical reflectivity, and provide both a nucleation layer for growth and a backside electrical contact for bottom-up optoelectronic structures. There have been reports of NW growth on metals such as Mo, Ti, Ta, and W [20-23]; however, the morphology of the NWs was shown to be affected by the instability of the metal layers during MBE growth [20]. On the contrary, ZrN buffer layers have been found stable during the metalorganic vapor-phase [24] and liquid phase [25] epitaxial growth of semiconductor structures. The high stability of ZrN was also confirmed by our recent study, in which we demonstrated vertically oriented Ga NW arrays grown by plasma-assisted MBE (PAMBE) on sputtered ZrN layers [26].

In this work, we show that the ZrN layer can serve as a high-quality backside ohmic contact for GaN NWs. We further provide a method that allows us to electrically characterize individual

NWs using nanomanipulators in the scanning electron microscope (SEM). The suitability of this approach is demonstrated by measuring the current-voltage (I-V) characteristics of individual GaN NWs with p-n junctions.

Electrical characterization of arrays of NWs is conventionally performed by depositing a large-area contact over an ensemble of NWs [16, 27-29]. A complex procedure of top electrode deposition includes the introduction of a polymer or glass filler between the NWs and multiple steps of top surface metallization [28]. In this configuration, the inhomogeneities between individual NWs result in varying resistances for each NW, causing parallel current injection that leads to uneven voltage drops across the NWs (the filament effect) [30]. To address the electrical properties of individual NWs, they can be transferred onto new insulating substrates with prefabricated electrodes [23, 31-36]. With this approach, however, the information about the as-prepared NW/substrate interface is inevitably lost. Here, we employ an alternative approach to characterize individual GaN NWs on the wafer using in situ I-V measurements within the chamber of a scanning electron microscope (SEM) [37, 38]. The method preserves the as-formed ZrN/NW interface and eliminates the technological steps required for the fabrication of electrical contacts, thus avoiding modification of the NW properties by fabrication processes. Moreover, the method can detect and quantify the dispersion of properties within the NW ensemble, such as variations in size, doping density, resistance, and location of the p-n junction.

**Materials and Methods**

**Sample preparation**

GaN NWs were grown by PAMBE in a RIBER Compact 21 system using an Addon radio frequency nitrogen source and a solid-source effusion Ga cell. C-plane sapphire substrates were covered with a 100 nm thick polycrystalline ZrN buffer layer using DC sputtering from a ZrN target. Layer deposition was performed at room temperature under an Ar pressure of $1.4 \cdot 10^{-3}$ mbar and a deposition rate of 0.13 nm/s.

The ZrN/Al$_2$O$_3$ substrates were transferred in air to the PAMBE system, heated to 150 °C for 1h in the loading chamber, and subsequently heated to 500 °C for 5h in the preparation chamber to remove any volatile contaminants before being transferred to the growth chamber. After heating the substrate to the required temperature, the PAMBE growth process was initiated by simultaneously opening Ga and N shutters. GaN NWs were grown under N-rich conditions with N and Ga impinging fluxes of $\Phi_N$ = 16 nm/min and $\Phi_{Ga}$ = 8 nm/min, respectively. Cross-sectional scanning electron microscopy (SEM) of thick GaN (0001) films grown under slightly N- and Ga-rich conditions at low temperatures (680 °C) was used to calibrate the Ga and N fluxes in the 2D-equivalent growth rate units of nm/min [39]. The stability of the N flux was controlled using an optical sensor of plasma light emission attached to the plasma cell [40]. During growth, the substrate was rotated at 6 rpm. The growth method was presented and discussed in detail in a previous publication [26].

N-type and p-type doping of GaN NWs were achieved using Si and Mg dopants, respectively. The doping level was modulated by adjusting the temperature of the dopant cell (T(Si) or T(Mg)). Three sets of samples grown at 818 °C were investigated: reference sample A contained homogeneously Si-doped GaN NWs with T(Si)=1055 °C and an expected doping concentration of $1 \times 10^{18}$ cm$^{-3}$; reference sample B consisted of homogeneously Mg-doped GaN NWs with T(Mg)=320 °C and an expected doping concentration of $2.5 \cdot 10^{18}$ cm$^{-3}$. Sample C is a GaN NW array with p-n junctions consisting of a 620 nm long GaN:Si segment and a 350 nm long GaN:Mg top segment. Both the GaN:Si and GaN:Mg segments were grown under the same conditions as the corresponding reference samples A and B. Table 1 lists the main parameters of the PAMBE growth and the properties of the GaN NWs. Figure 1 schematically represents the structures investigated in this study, including the different configurations of the metal contacts employed for electrical characterization. The backside contact for all NWs was provided by the ZrN buffer layer. The top contact was formed either by a tungsten (W) tip in direct contact with the bare top facet of the NW, or by sputtered metal layers to establish ohmic contacts. The parameters and purpose of the additional metallization are discussed below.

**Methods of characterization**

The morphology and dimensions of the as-grown GaN NWs were analyzed by SEM (Tescan Lyra3). The I-V characteristics also were measured in the chamber of the Tescan Lyra3 dual-beam microscope using nanomanipulators designed for low electrical current measurements. A DC bias was applied to the tungsten tip of the nanomanipulator (OmniProbe 400, Oxford Instruments) that contacted the top segment of the individual NW, while the bottom ZrN contact was grounded.

Figure 2a shows the schematic of the measurement setup. Before the I-V measurements, the tip of the nanoprobe was shaped using a focused Ga$^+$ ion beam (FIB). FIB-assisted shaping of the nanoprobe has proven to be an appropriate approach for reducing the contact resistance by removing the native tungsten oxide from the nanoprobe tip [38]. By contacting individual NWs in a dense array, the leakage current can be eliminated in contrast with large-area contacts of

NWs, where the close contacts between neighboring NWs may form conductive paths bypassing the p-n junctions.

Figure 2b shows the SEM image of the nanoprobe in contact with the GaN NW. Nanomanipulators in the chamber of the SEM allow us to characterize an individual NW of choice and simultaneously control the reliability of the contact.

## Results and discussion
### Morphology of NW arrays

Figure 3 shows a cross-sectional view of the GaN:Si NWs (sample A), GaN:Mg NWs (sample B), and NWs with an axial p-n junction (sample C) grown by PAMBE on a ZrN/sapphire substrate. All the NWs are vertically oriented despite the expected epitaxial link of GaN to randomly oriented ZrN grains, similar to that reported previously for the self-induced growth of GaN NWs on TiN [41]. We explained this finding by the effect of geometrical selection at the initial stage of growth, during which the unidirectional supply of species in PAMBE favors the vertical growth of the NWs [26]. A slight inverse tapering was observed on sample B and on Mg doped segments of sample C. Such widening of the NWs in their top parts results from an enhancement in the radial growth induced by Mg incorporation during the growth of Mg-doped GaN NWs [42-44].

### I-V measurements

First, the ZrN buffer layer was characterized by measuring the I-V characteristics of the W tip-ZrN-W tip structure. Figure 4 shows a linear I-V characteristic with a low resistance of 8 Ω, extracted from the slope of the curve. This indicates that the metallic ZrN layer contributes negligibly to the electrical properties of ZrN/GaN structures.

Next, ZrN/ GaN:Si NW structures were investigated using a bare W tip as the top contact to the NW. The I-V characteristic of the ZrN/ GaN:Si/W structure (Figure 5a) exhibited diode-like behavior with a turn-on voltage of 1 V, rectification ratio of $10^5$ at ±2 V, and series resistance $R_s = 20$ kΩ. The forward-bias branch of the I-V curve under a positive voltage applied to the top contact indicates the existence of a Schottky barrier at the top contact.

Numerical simulations of the forward bias I-V characteristics using a conventional TE model in MATLAB were performed to estimate the parameters of the Schottky contact. According to Rhoderick [45], the current density in TE theory is given by

$$j = j_0 exp\left(\frac{eV_A}{\eta k_B T}\right)\left[1 - exp\left(-\frac{eV_A}{k_B T}\right)\right], \qquad (1)$$

where reverse bias saturation current is

$$j_0 = A^* T^2 exp\left(-\frac{\Phi_B^{eff}}{k_B T}\right). \tag{2}$$

Here e is the elementary charge, $k_B$ is the Boltzmann constant, A* is the Richardson constant corresponding to the effective mass in the semiconductor, T is the absolute temperature, VA is the applied voltage, $\Phi_B^{eff}$ is the effective barrier height, and $\eta$ is the ideality factor.

For biases $V > 3\,k_B T/e$ and in the presence of series resistance Rs and shunt resistance Rp, Equation (1) can be rewritten as:

$$j_{SB} = j_{0,SB}\left[exp\left(-\frac{e(V-iA_0 R_s)}{\eta k_B T}\right) - 1\right] + \frac{V - iA_0 R_s}{R_p} \tag{3}$$

The values obtained by fitting the experimental data with formula (3) are series resistance $R_s$=60·10$^3$ Ω, shunt resistance $R_p$=10$^{10}$ Ω, saturation current $I_0$=1·10$^{-14}$ A, and ideality factor $\eta$=1.75.

In the next step, a TiAl metal layer (10+10 nm) was thermally evaporated on top of the GaN NWs to eliminate the Schottky barrier at the W/GaN interface [46]. The sample was tilted by 45° to limit the metal deposition to the top segment of the NWs and to avoid metallization of the NW sidewalls, as schematically shown in Figure 1. The linear I-V characteristics obtained after the deposition of the top TiAl contact confirm the ohmic behavior at both the bottom ZrN/ GaN:Si and top GaN:Si/TiAl interfaces (Figure 5b).

The resistance of the GaN:Si NW extracted from the slope of the linear I-V characteristic curve was RNW=450 Ω. The carrier mobility can be calculated as μ=1⁄qρn, where q is the elementary charge, ρ is the resistivity, and n is the charge carrier concentration. Assuming that the diameter of the NW is 0.4 μm, the length of the NW segment not covered by the metal is 1 μm, and the carrier concentration is equal to the doping level, we obtain an electron mobility of 1·10$^3$ cm$^2$·V$^{-1}$·s$^{-1}$, which is a reasonable value for high-quality GaN NWs [47].

Concluding this part, we demonstrated that the ZrN buffer layer forms an ohmic contact with n-type GaN NWs and that the bare tungsten nanoprobe is unsuitable for characterizing Si-doped GaN NW, as it forms a nonlinear contact with GaN.

Next, the ZrN/ GaN:Mg NW structures were investigated using a bare W tip as the top contact to the NW. The symmetrical nonlinear I-V characteristic (Figure 6a) can be attributed to two back-to-back Schottky diodes [48] formed on both the top and bottom interfaces of a single GaN:Mg NW. To eliminate the top Schottky contact, a NiAu metal layer was deposited to form an ohmic top contact with GaN:Mg [46]. Figure 6b shows the I-V characteristic after metallization. The diode-like shape of the I-V curve with forward bias under a positive voltage applied to the top contact indicates an ohmic GaN:Mg/NiAu interface and Schottky ZrN/GaN:Mg interface.

Figure 7 shows the I-V characteristic of the ZrN/p-n GaN NW structure. A NiAu contact was deposited on top of the GaN:Mg segment of the NW to prevent the formation of a Schottky barrier between the W tip and GaN:Mg. The I-V characteristic exhibited pronounced rectifying behavior, with a turn-on voltage of 2 V and a rectification ratio of $4\cdot10^4$ at ±2 V. Assuming the ohmic nature of both the top and bottom metal contacts, we attribute the rectifying I-V characteristics to the p-n junction embedded within a single GaN NW. The measured reverse bias current was significantly higher than that predicted by the theoretical model used in this study. We attribute this effect to parasitic current leakage under reverse bias [49, 50]. The pronounced leakage may be caused by the parasitic radial p-type shell overgrowing the p-n interface, resulting in the mixed axial and radial behavior of the NW-based p-n junction [51]. To estimate the parameters of the p-n junction formed inside the GaN NW, we performed numerical simulations of the forward bias I-V characteristics in the frame of a TE model (Equations 1 -3). Solving Equation (3), we extracted the following values of the parameters of the p-n junction: series resistance $R_s=40\cdot10^3$ Ω, shunt resistance $R_p=\cdot10^{14}$ Ω, saturation current $I_0=1\cdot10^{-10}$ A, and ideality factor $\eta=4$.

Larger ideality factors ($\eta>2$ for p-n heterojunctions) can be accounted for by an inhomogeneous barrier distribution, particularly when considering the local series resistances [52, 53]. Such inhomogeneity is likely induced by the structural disorder and defects close to the p–n interface [54]. The structural disorder and defects are located on p-type side of the p-n junction as it has been reported already that Mg incorporation might be responsible for the degradation of the GaN quality [27]. Moreover, it is known that Mg incorporation in GaN NWs is highly nonuniform across the NW lateral dimensions [44], with a tendency to segregate at the NW side surfaces [42]. However, the diode performance of the single NW p-n junction presented in this study shows a significant improvement over previously reported values for an ideality factor $\eta>10$ [16, 34], turn-on voltage $V_{on}=3.5 – 6$ V [23, 32, 36], leakage current 90 fA – 80 nA [28, 33], and series resistance $10^2$-$10^4$ MΩ [32].

**Conclusions**

Polycrystalline ZrN layers have been demonstrated to provide backside ohmic contact to dense vertical arrays of n-type GaN NWs grown by PAMBE, which is of particular significance for their application in optoelectronic devices. Evidence for the achievement of an ohmic contact was obtained by measuring the electrical characteristics of the as-grown single NWs on

ZrN buffer layers using nanomanipulators in SEM. Measurements employing nanomanipulators facilitated the determination of the key electrical properties of individual semiconductor NWs, including electron mobility and charge carrier concentration. As a proof of concept, single NWs with a p-n junction were characterized, exhibiting highly rectifying characteristics. Numerical analysis of the current-voltage characteristics indicated that charge transport in the forward direction was governed by the thermionic emission mechanism.

**Acknowledgments**

The authors are grateful to M. Guziewicz from Łukasiewicz Research Network - Institute of Photonics and Microelectronics in Warszawa for deposition of ZrN buffer layers on sapphire substrates. This research was partly funded by the Polish National Science Centre NCN grants 2021/43/D/ST7/01936 and 2022/04/Y/ST7/00043 (Weave-Unisono) and by the Czech Science Foundation project 23-07585K.

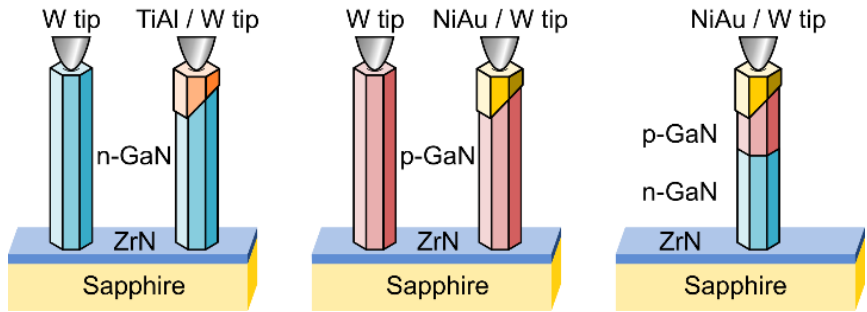

**Figure 1.** NW-based structures with different configurations of metal contacts. Three types of GaN NWs grown by PA MBE were characterized: n-type GaN:Si NWs, p-type GaN:Mg NWs, and GaN NWs with axial p-n junctions. The bottom contact for all NWs was provided by the ZrN buffer layer. The top contact was formed either by a bare tungsten (W) tip or by sputtered metal layers (TiAl for GaN:Si and NiAu for GaN:Mg) to establish an ohmic contact between the W tip and the NW.

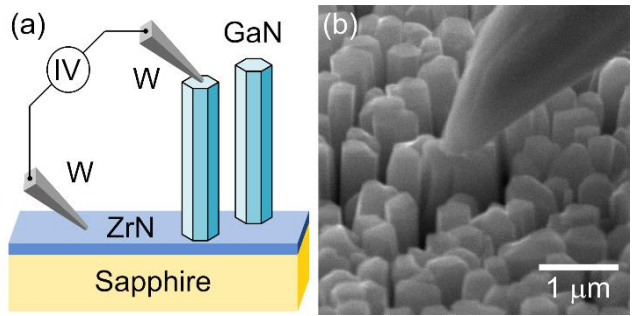

**Figure 2.** (a) Measurement setup for nanomanipulator-based electrical characterization of as-grown individual GaN NWs on the ZrN buffer layer within the SEM chamber. (b) SEM image of the tungsten tip of the nanomanipulator in contact with an individual GaN NW. The tip was shaped by focused Ga+ ion beam prior to electrical measurements.

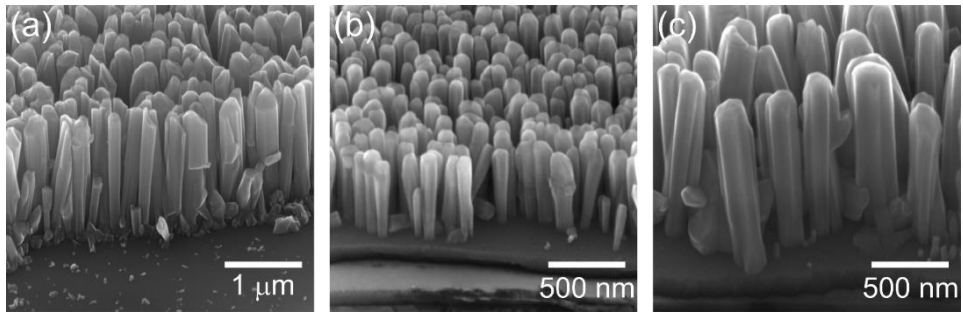

**Figure 3.** Bird-view SEM images of GaN NWs grown on ZrN substrates: (a) GaN:Si NWs, (b) GaN:Mg NWs, and (c) NWs with axial p-n junctions.

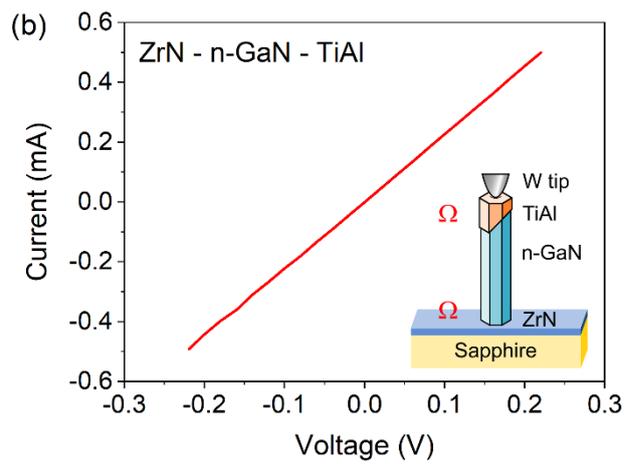

**Figure 4.** I-V characteristic of the ZrN layer measured using two nanomanipulators with W tips in the SEM chamber; inset: schematic illustration of the structure.

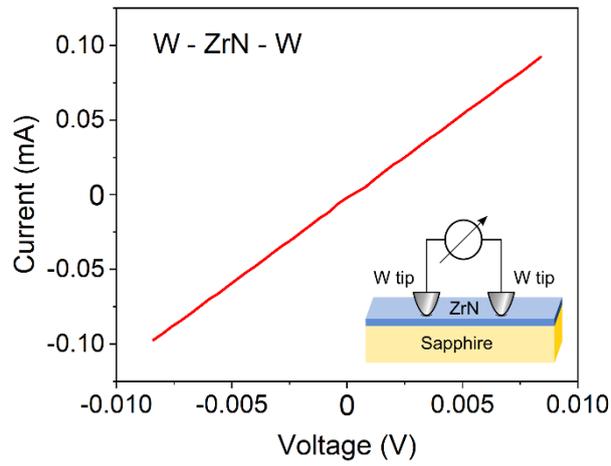

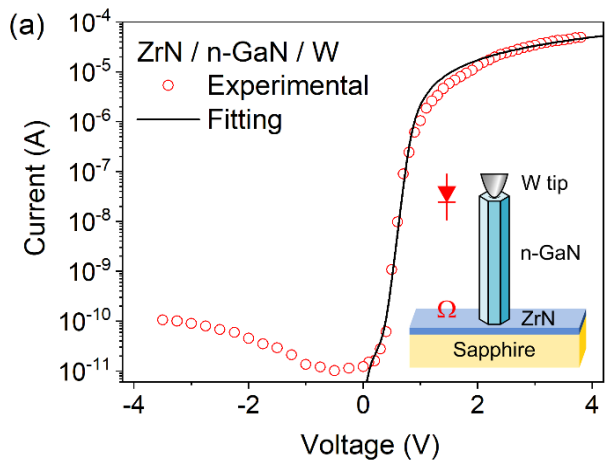

**Figure 5.** SEM I-V characteristics of ZrN/ GaN:Si/W (a) and ZrN/ GaN:Si/TiAl (b) structures; inset: schematic illustration of the structures; red symbols highlight respective ohmic and rectifying interfaces.

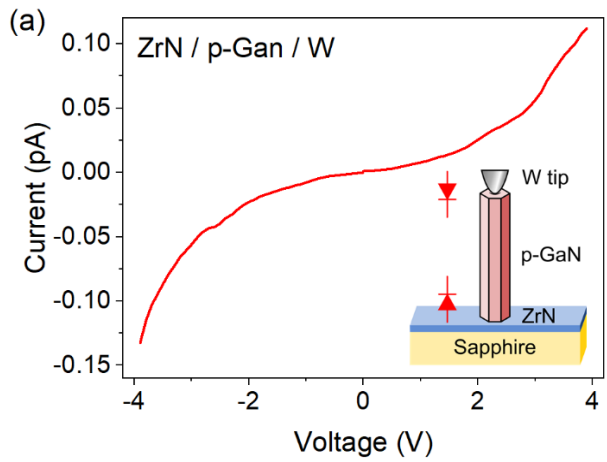

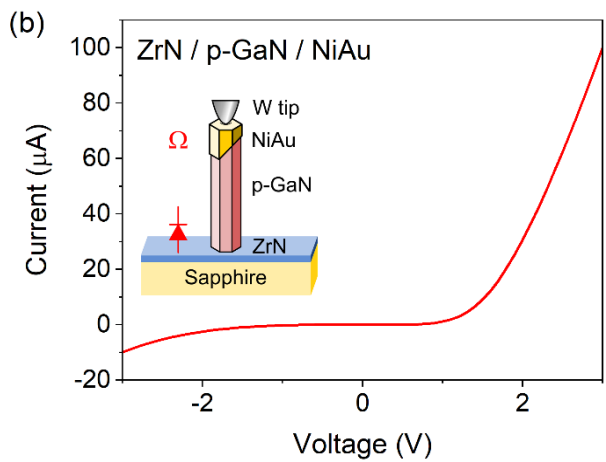

Figure 6. SEM I-V characteristics of ZrN/ GaN:Mg/W (a) and ZrN/ GaN:Mg/NiAu (b) structures; insets: schematic illustration of the structures; red symbols highlight respective ohmic and rectifying interfaces.

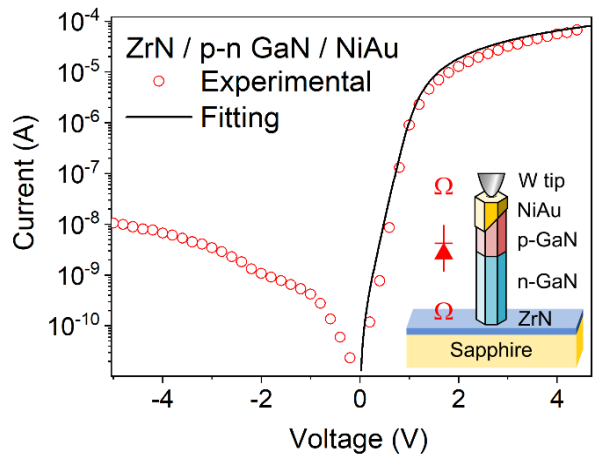

Figure 7. Experimental (symbols) and numerically fitted (solid line) SEM I-V characteristics of the ZrN/p-n GaN/NiAu structure; inset: schematic illustration of the structure; red symbols highlight respective ohmic and rectifying interfaces.

**Table 1.** Deposition parameters for PAMBE growth and properties of the resulting NWs.

| Sample # | type | NWs length/width | Dopant cell temperature | Doping level | Top metal contacts |
|---|---|---|---|---|---|
| A | n-type | 2 / 0.2-0.4 µm | 1055°C, Si | $1 \cdot 10^{18}$ cm$^{-3}$ | tungsten (W) tip or TiAl (10+10nm) |
| B | p-type | 1 / 0.2 µm | 320°C, Mg | $2.5 \cdot 10^{18}$ cm$^{-3}$ | tungsten (W) tip or NiAu (10+10nm) |
| C | p-n junction | 1 / 0.2 µm | 1055°C, Si 320°C, Mg | (n) $1 \cdot 10^{18}$ cm$^{-3}$ (p) $2.5 \cdot 10^{18}$ cm$^{-3}$ | NiAu (10+10nm) |